\documentclass{mn2e}
\usepackage{epsf}
\usepackage{graphicx}

\def\etal{{\it et~al. }}

\title[Millimetre observations of Circinus X-1]
{Millimetre observations of a sub-arcsecond jet from Circinus X-1}

\author[D.E. Calvelo \etal]
{D.E. Calvelo,$^{1}$ R.P. Fender,$^{1}$ A.K. Tzioumis,$^{2}$ J.W. Broderick $^{1}$ \\
$^{1}$ School of Physics and Astronomy, University of Southampton,
Highfield, Southampton, SO17 1BJ, UK\\
$^{2}$ Australia Telescope National Facility, CSIRO, PO Box 76, Epping, New south Wales 1710, Australia\\}

\date{\today}

\volume{000}

\pagerange{\pageref{firstpage}--\pageref{lastpage}} \pubyear{2010}

\begin{document}

\label{firstpage}

\maketitle
\begin{abstract}

We present results from the first successful millimetre (combined 33 GHz and 35 GHz) observations of the neutron star X-ray binary Circinus X-1, using the Australia Telescope Compact Array. The source was clearly detected in all three observing epochs. We see strong evidence for a periastron flare beginning at MJD 55519.9 $\pm$ 0.04 with estimated peak flux densities of up to 50 mJy and which proceeds to decline over the following four days. We directly resolve jet structures on sub-arcsecond scales. Flux density variability and distance from the core of nearby components suggests recent shock re-energisation, though we are unable to directly connect this with the observed flare. We suggest that, if the emission is powered by an unseen outflow, then a phase delay exists between flare onset and subsequent brightening of nearby components, with flows reaching mildly relativistic velocities. Given resolved structure positions, in comparison to past observations of Cir X-1, we find evidence that jet direction may vary with distance from the core, or the source's precession parameters have changed.

\end{abstract}


\section{Introduction}

Circinus X-1 (Cir X-1) is a confirmed (Linares \etal 2010) neutron star X-ray binary (NSXRB) known for its regular 16.6 day periastron flares (radio: Whelan \etal 1977, IR: Glass 1978, X-ray: Tennant, Fabian \& Shafer 1986), and for being one of the few NSXRBs to show relativistic jets, resolved at a variety of scales and wavelengths (X-ray: Heinz \etal 2007; Soleri \etal 2009, radio: Stewart \etal 1993; Fender \etal 1998; Tudose \etal 2006). In the past, Cir X-1's flares were found to precede brightening of nearby ejecta, which was interpreted as re-energisation via interaction with unseen outflows. The time delay between radio core flaring and re-brightening of the downstream material indicated Lorentz factors of $\Gamma$ $>$ 15 (Fender \etal 2004), and while this has been corroborated via further advanced analysis of the same data-sets (Tudose \etal 2008, henceforth Tu08), subsequent observations of the source have yet to yield similar results (Calvelo \etal 2011 - in press, henceforth Ca11).

Currently, there remains uncertainty on the orientation of the system and its jets. The ultra-relativistic flow velocities would imply a jet inclination very close to the line of sight ($\theta$ $<$ 5$^{\circ}$: Fender \etal 2004). However, fitting of blue and red-shifted X-ray emission features (possibly originating in the jets) have been used to calculate lower jet speeds (0.08c) and a jet inclination near perpendicular to the line of sight (Iaria \etal 2008). Furthermore, there exist multiple pieces of evidence for an edge-on accretion disc, including X-ray P-Cygni profiles (Brandt \& Schulz 2000) and X-ray dips (Shirley, Levine, \& Bradt 1999). These results need not be contradictory, if we assume that the jets are either misaligned with the orbital plane (e.g. Maccarone 2002), follow significantly non-linear flow paths (already indicated on larger scales in radio maps; Tudose \etal 2006), or precess. 

Tu08 reported little variation in the jet axis over a decade of observations (1996-2006) implying minimal precession in the system over long periods of time. In contrast, more recent observations (2009/2010: Ca11) suggest that the source's 3-dimensional jet orientation may now differ from that observed over the 1996-2006 period (projected angle difference of 40 $\pm$ 5$^{\circ}$ and further from the line of sight). Ca11 also found that the most significant structural variability appeared to the north-west of the source (previously a region of minimal activity and speculated to be the direction of the receding jet: Tu08), as well as structures within the nebula reminiscent of the filaments observed in nebula W50, surrounding SS 433 (Dubner \etal 1998). Ca11 postulate that the system's jets may have shifted recently, or even precess in a similar manner to those of SS 433.

\begin{table*}
\caption{Cir X-1 observations. The table lists the date, start Modified Julian Day (MJD) of each observation (on-source), total on-source time before and after flagging, MJD end time after flagging, daily measured peak flux densities (33-35 GHz) for the core and north-western component, and noise levels of the images from which the measurements were taken.}
\begin{center}
\begin{tabular}{|l|c|c|c|c|c|c|c|c|}
Date & MJD & On-source & Post-flag & Post-flag & Core & Component N & rms \\
(UT) & start & time (h) & on-source time (h) & MJD end & (mJy beam$^{-1}$) & (mJy beam$^{-1}$) & (mJy beam$^{-1}$) \\
\hline
2010 Nov 19 & 55519.841 & 8.86 & 6.08 & 55520.375 & 7.14 & 2.11 & 0.23 \\
2010 Nov 21 & 55521.836 & 8.42 & 3.31 & 55522.021 & 1.66 & 0.52 & 0.08 \\
2010 Nov 23 & 55523.804 & 8.14 & 5.47 & 55524.240 & 0.25 & 0.10 & 0.02 \\
\hline
\end{tabular}
\vspace{-10pt}
\end{center}
\end{table*}

The system's quiescent and flare levels have varied significantly since discovery, with radio flares reaching above 1 Jy in the late 1970s (Haynes \etal 1978), but declining since 1997 to reach only 10s of mJy (Fender, Tzioumis, \& Tudose 2005). Though inter-flare monitoring indicates Cir X-1 remains historically `faint', recent activity (X-ray flares greater than 100 mCrab: Nakajima \etal 2010, radio flares greater than 0.1 Jy: Calvelo \etal 2010) has renewed observational interest in the object. Based on recent observations with the Australia Telescope Compact Array Broadband Backend (ATCA-CABB) in Ca11, spectral index estimates indicated Cir X-1 would likely be detectable at mm wavelengths. NSXRBs had yet to be observed successfully at this frequency; thus, a detection would not only fill an unexplored spectral region but also allow for higher resolution images than had previously been achieved with ATCA. 

\section{Observations \& Data Reduction}

Observations were carried out on 2010 Nov 19-20, 21-22 and 23-24 using the ATCA-CABB in 6A configuration (minimum baseline of 337m, maximum of 5939m) at both 33 and 35 GHz (details in Table 1). PKS B1253-055, PKS B1934-638, and PKS B1511-55 were used as the bandpass, flux and phase calibrators respectively. Between 8 and 9 hours of time was spent on-source each day (predicted rms noise of 26 - 28 $\mu$Jy at both frequencies - average weather conditions). All data and image processing was carried out in {\sevensize MIRIAD} (Sault, Teuben and Wright 1995).

Since millimetre observations are more susceptible to atmospheric effects than longer wavelengths, phase stability had to be closely monitored via the ATCA ``seeing'' monitor, which displays rms path length fluctuations over 10 minute samples in microns (Middelberg \etal 2006). Phase stability declined as runs progressed, due to rising temperature and humidity. Though a switch to lower frequencies was never warranted (rms path length fluctuations tended to remain below 700 microns), once analysis began it was found that large segments towards the latter half of each observation did not have sufficient phase stability to be reliable.

Due to phase stability decay, it was necessary to define the level of flagging that would improve final image fidelity. By dividing observations into segments, and comparing the resulting maps to rms path length fluctuations ({\sevensize VARPLT}), we found that segments with low rms showed the expected point source at Cir X-1's location. Higher rms segments corresponded with increasing noise towards Cir X-1's position, with no single distinguishable point source. In general, segments whose majority of time was spent with rms higher than 400 microns failed to produce the defined structure. Thus for the remainder of analysis, we only use data whose rms values remained below 400 microns.

\section{Analysis \& Results}

To produce maps with high signal to noise ratios, we combined 33 and 35 GHz data during the inverse Fourier transform of calibrated uv data (i.e. included both frequency's visibility files with multi-frequency synthesis in {\sevensize MIRIAD}'s {\sevensize INVERT} step). A complete spectral analysis will be covered in a future publication.

\subsection{Light Curves}

Declining phase stability prevented production of complete sets of image plane measurements for each day's full observation. However, having produced maps for each day, it is apparent that Cir X-1 is the only visible source, and thus we may estimate flux density variability from {\em uv} data amplitude plots. We are, however, still unable to distinguish whether variation arises from the `core' of Cir X-1 (designated as a point source situated at Cir X-1's well established co-ordinates: Tu08) or surrounding structure (section 3.2). 

Figure 1 shows each observation's visibility amplitudes, with available image plane fits over-plotted. The most noticeable feature is the multi-peaked rise in the first run's amplitudes, which in comparison to the two subsequent runs, suggests flaring activity. A rise is also noticeable in the image plane fits, though it only includes a few points for each run. The multiple peaks may be attributed to real variations from Cir X-1 (multi-peaked flares have been seen before: Thomas \etal 1978, Tu08), or resolved structure. In this case the radio data alone is insufficient to eliminate either possibility.
Our inability to produce high fidelity maps for the time periods of highest amplitude means we cannot accurately measure the peak flux density reached by this flare, however, if we assume the relative difference between mean average amplitude and core image plane values remains similar throughout the first observation, we may crudely estimate that flux densities could have reached up to 50 mJy.

\begin{figure}
\centerline{\includegraphics[width=3.3in]{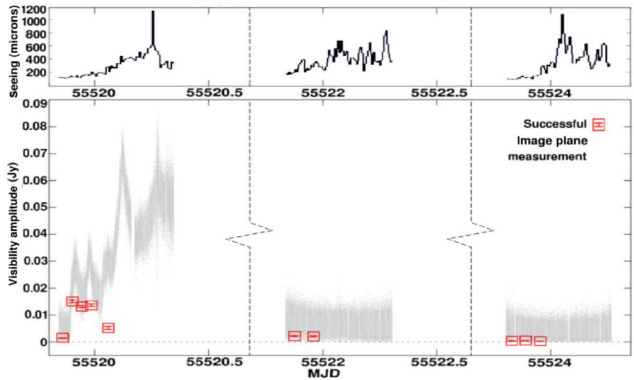}}
\caption{Circinus X-1 ATCA-CABB rms path length fluctuations (top), {\em uv} amplitude, and image plane fit plots (three observations over 5 days). Measurement errors (1$\sigma$) are included in the image plane values but are too small to appear clearly.}
\vspace{-10pt}
\end{figure}

The second run's amplitudes indicate a post-flare decline, although the two image plane measurements show a slight rise. With this observation data set having been the most affected by phase disturbances, it is difficult to support either claim. The final observation set shows the lowest amplitudes and flux densities, with the source having likely returned to quiescent levels. The flux densities measured from the core on each individual day are shown in Table 1 and also show the decline in flux across the three observations, supporting much of what is observed in the plots.

\subsection{Imaging}

Individual image contour plots for each day are shown in Figure 2. Cir X-1 is resolved in all three runs, at a position that shows no shift greater than the positional errors associated with each set of images. Jet like structure is visible along a single axis (north-west to south-east) which is similar to that seen in previous mapping of the source at cm wavelengths (Tudose \etal 2006, 2008; Ca11). 

When interpreting the images, we must remain aware of the fact that we have imaged the source in both a state of rise and of decay, violating one of the major assumptions in aperture synthesis. As a result, artefacts may have been introduced into the image which can imitate the appearance of symmetrical jets. Simulations were created based on the visibilities from the first observation where we assumed all emission was from a single point source at Cir X-1's location with flux density S$_{\nu}$ $\approx$ average visibility amplitude. Images produced from the simulated data did indeed show signs of axial `spokes' from the central source during peaks in amplitude; however, the intensity of these spokes beyond 0.5 arc-seconds remained below 10 per cent of the core flux density (i.e. lower than the 3$\sigma$ noise levels measured in our images) and the more intense residuals left within 0.5 arc-seconds could be identified by their symmetrical layout and pairing with nearby negative regions of similar intensity.

While we acknowledge that our first epoch image must be affected by the above artefacts to some degree, we remain convinced that much of the resolved structure cannot be a result of this for a number reasons. Firstly, as per the simulations, the artefacts should appear symmetrical about the point of variation, and while there is a degree of symmetry in the visible components, two distinct asymmetric features arise repeatedly in all images. These `extensions' to the south-east and north-west of the core appear to differ in both structure and intensity, and a separate emission component appears 0.75 $\pm$ 0.25  arc-seconds to the north-west of the core (slightly below the `jet' axis). Secondly, the flagged data sets all share a similar starting hour angle, and though the second flagged data set is significantly shorter than the others, the position angle of the resulting beams (-30.7$^{\circ}$, -40.9$^{\circ}$, -41.1$^{\circ}$ for Nov 19, 21 and 23 respectively) remains similar across all data-sets. With this in mind, we would expect the axis of visible artefacts to differ between the first and second runs, as we are dealing with opposite types of light curve behaviour - rise and decay - and yet they remain similar. Thirdly, the final data set shows little or no evidence for continuing flux density decay, and as such should be free of the above effects. Yet, we continue to observe elongation of the core and the north-component in the final day's images.

There is a further concern in that data whose level of phase errors gradually vary could produce similar artefacts to those from source variability. Though we have taken steps to eliminate afflicted data, this does not mean our images are not affected to some degree (see Taylor, Carilli \& Perley 1999, chapter 28). The random nature of these errors means that over a large enough integration, artefacts should still appear symmetrical, making the first argument of the previous paragraph, coupled with multiple detections, our strongest test for validity; i.e. it is unlikely that persistent asymmetrical structure will be an artefact.

\begin{figure}
\centerline{\includegraphics[width=3.1in]{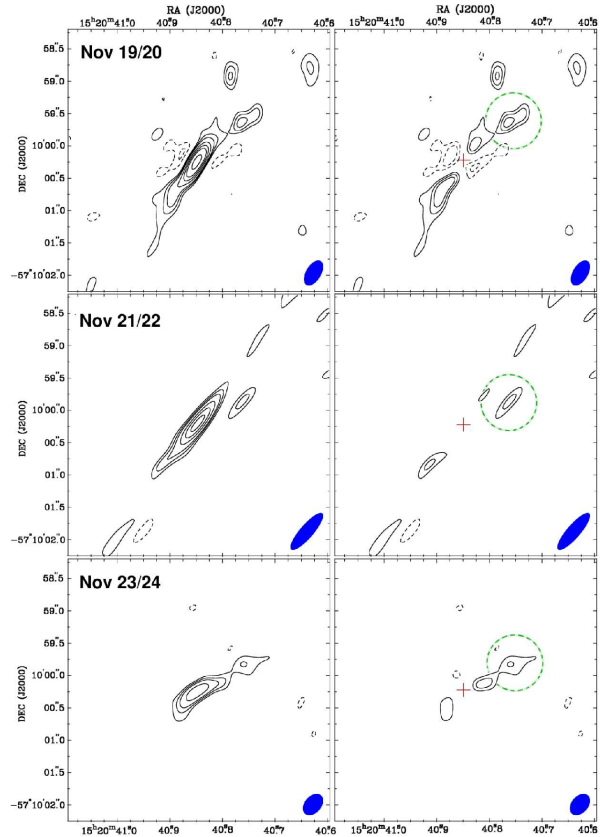}}
\caption{Radio contour maps of Circinus X-1, created using combined 33 and 35 GHz data. Weighting uses a robust factor of 0.5 (an optimal compromise between natural and uniform weighting). Contours are set at  -6,-4.5,-3,3,4.5,6,9,12,15,21,27,36 $\times$ rms noise of each epoch. Beam sizes are 0.44$\times$0.24, 0.76$\times$0.21, and 0.39$\times$0.26 arcsec for the 19th, 21st and 23rd respectively. The right panels have had a fitted point source subtracted at the core's position (the cross). The dashed circle marks component N.}
\vspace{-10pt}
\end{figure}

To further examine resolved structure we performed point source subtraction centred at the core, with the results shown in Figure 2 (right hand panels). The map of the 19th (first epoch) shows significant residual structure, but since it is the day of the flare, artefacts caused by source variability are more likely to occur. Signs of such effects appear in proximity to the subtraction point as roughly symmetrical negative zones along an axis perpendicular to that of positive structure. Even assuming the presence of artefacts, they cannot account for the brighter south-eastern core residual (approximately 50 per cent brighter; a difference of 0.8 mJy) nor the three other separate emission components in the north west quarter. Out of these remaining bright components, only the closest to the core (marked by the dashed circle in Figure 2, henceforth referred to as component N) remains in all subsequent images. The two structures farther from the core may be real and simply fade between the first and second observations (perhaps related to an earlier ejection event) but without multiple detections this is difficult to confirm.

The second map set also shows both south-eastern residuals and component N. We observe a streak artefact running the length of the image, parallel to the apparent jet axis, which we were unable to eliminate. This appears to be the source of several weak structures across the length of the axis, and also affects the southern extension. 

The final observation continues to show component N and faint extension residuals; however, the northern extension now appears slightly offset from the `jet' axis (further south than earlier examples) and is the more dominant of the pair of core residuals, though the difference in flux density is sufficiently low for noise to account for it. With variability likely to have reached minimum levels (as indicated by the image plane measurements), if any artefacts now exist in the map they should be a result of phase errors. This cannot be ruled out a cause of the low level extensions near the core.
 
\begin{figure}
\centerline{\includegraphics[width=3.1in]{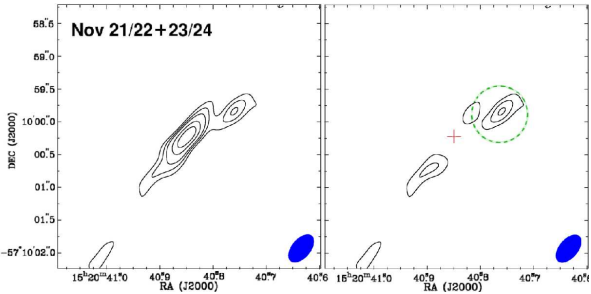}}
\caption{Radio contour map of Circinus X-1, created using combined data from 2010 Nov 21 and 23 (33 and 35 GHz). Natural weighting is used for higher sensitivity at the cost of slight reduction in beam resolution (beam size = 0.50$\times$0.29 arcsec). Contours are set at  -6,-4.5,-3,3,4.5,6,9,12,15,21,27,36 $\times$ rms noise of 53 $\mu$Jy. Layout matches that used in Figure 2.}
\vspace{-10pt}
\end{figure}

Figure 3 shows naturally weighted images from combined 21st and 23rd observations (post flare), resulting in a greater than 6$\sigma$ detection of component N. Accepting component N as a real structure, it may be moving ejecta from Cir X-1 or a shock caused by re-energisation of slow material by an unseen flow (as in Fender \etal 2004). We find that its flux density declines over the observations (see Table 1), though we cannot determine the onset of a rise as the object is difficult to detect on shorter data intervals. LBA observations in our partnered submission, Miller-Jones \etal (submitted to MNRAS, henceforth MJ2011), suggest the presence of objects with proper motions of 35 mas day$^{-1}$. Ejecta moving at such speeds in our observations would result in detectable shifts of position between our initial and final images. Analysis of component N's position in our three images shows a 0.29 $\pm$ 0.13 arc-second shift, corresponding to a proper motion of about 150 $\pm$ 70 mas d$^{-1}$ to the south between the first and second epochs. This is the only significant change as all other shifts fall below respective fitting error limits. The shift actually brings the component closer to the core, perhaps implying a bent outflow or a sequence of different fading and brightening jet components. However, being a marginal difference and including the possibility that image errors have affected the component's position (i.e. more significant in the first day image) makes it difficult to justify further analysis.

Furthermore, based on revised ephemeris estimates for the system (George Nicolson, private communication) our epoch's flare should have occurred at MJD 55519.94; about an hour after the rise in the amplitude plot begins. Thus, component N's distance from the core prohibits association with our flare, as it would imply a proper motion over 2000 mas d$^{-1}$. This argument would also be true for re-energisation via internal shocks. Such constraints, however, do not apply if the event is connected to an earlier flare. Comparison to observations of mm flares from black hole (BH) XRB GRS 1915+105 (Fender \& Pooley 2000) suggests that ejecta emission declines rapidly at mm wavelengths (GRS 1915+105 events have duration shorter than 1 hour), making an ejected component's possible life time of over 15 days questionable. Even if we assume a shallower decay, extrapolation of component N's decay rate back towards the earlier flare would imply minimum flux densities greater than 50 mJy at launch; comparable to the peak core flare levels estimated from the amplitude plots. Shocks do not suffer such limitations, as the energising flows may travel undetected before finally brightening on interaction with slower media downstream (as is observed in other BHXRBs, e.g. XTE J1550-564: Corbel \etal 2002 and NSXRBs, e.g. Scorpius X-1: Fomalont, Geldzahler, and Bradshaw 2001). Thus, if we assume a single orbit's delay (16.6 days) in component N's variability, we estimate proper motion for a flow of approximately 60 mas d$^{-1}$.

One may also be tempted to carry out similar calculations based on the slight shifts in position observed in the southern and northern extensions, but the possibility that they are affected by any of the artefacts we have discussed would severely limit our confidence in such an exercise.

\section{Discussion}

Having established that image artefacts predominantly affect the structures nearest to the core, our best jet axis estimate uses component N's position, giving an angle of 121 $\pm$ 5$^{\circ}$ E of N. Though this estimate falls within the range quoted in Tu08 (129 $\pm$ 13$^{\circ}$ E of N) and is only marginally steeper than the estimate from milli-arcsecond structure in MJ2011's LBA observations (113$^{\circ}$ E of N), it is approximately 50$^{\circ}$ shallower than the axis estimated from 5.5 GHz residual emission near the core in Ca11. Furthermore, whereas MJ2011 observe symmetrical jets, our maps indicate a scenario much like that observed in Ca11 and Tu08 where asymmetry of near-core residuals indicates a brighter southern jet. Like Ca11, we also observe distinct activity from a region that has previously been associated with the receding jet (Tu08) in the form of component N.

The fact that the various structural orientations appear on different scales across observations taken within months of each other supports a scenario where the jet angle varies as we move further from the system's core. Starting off at relatively high angles to the line of sight and near E-W as indicated in MJ2011, but steadily turning N-S and closer to the line of sight as we move down the flow to arcsecond scales as in Ca11. The origin of such deviation may be precession, or quasi-static jet kinks. The unexpected appearance of structures to the north west (component N in this case) may be a result of asymmetry in the density of interacting media, or deviations in a jet close to the line of sight (as suggested in Ca11).

The jet velocities reported in MJ2011 are of similar order to our estimates for energising flows in section 3 (when we assume a full orbit's phase delay), both of which fall far below the ultra-relativistic speeds calculated by Fender \etal (2004). However, we too cannot completely rule out the existence of such ultra-relativistic flows. There remains the possibility that we did not observe the full flare event associated with our specific orbit (ephemeris estimates are not perfect, nor every flare event the same), altering the reference time on which our component N calculation is based.

Reconciling these combined results with historic observations such as the ultra-relativistic velocities of Fender \etal (2004) and lack of precession observed in Tu08 is difficult. Lacking details on sub-arcsecond structure at the time of the Tu08 observations, we cannot establish whether Cir X-1's jets have always exhibited variable inclination with distance from the core (though jet curvature in large scale images from Tudose \etal 2006 indicate some level of this). By grouping radio observations made after 2008 and comparing them to older results, we begin to observe an almost bimodal divide in Cir X-1's behaviour, with lower jet velocities, precession (indicated by the difference in jet angle observed at 5.5 GHz: Ca11) and multiple incidences of activity to the north-west of the core rather than the south-east being recent changes. This may indicate that, along with the decline in intensity over the last decade, the system's jet structure can also vary significantly with time.

Such a scenario draws further similarities between Cir X-1 and SS433/W50. Hydrodynamical simulations by Goodall \etal (2011) show that the structures visible in nebula W50 cannot be reproduced by extrapolation of SS433's current jet activity back through time. Instead they show that at least three distinct epochs of jet activity, each with a different level of precession, were required to produce the nebula's layout. If the filaments visible in Ca11's Cir X-1 nebula maps are indeed caused by jet-ISM interactions, then the precession angle of the jets would have had to be significantly larger than that indicated by pre-2008 radio observations. Thus, variability of Cir X-1's jet characteristics should not be entirely unexpected.

\section{Summary}

We have detected Circinus X-1 and sub-arcsecond structure at millimetre wavelengths for the first time. Light curves indicate the rise phase of a periastron flare event occurred on Nov 19, decaying over the following four days. Radio maps show resolved structure near the core and a 6$\sigma$ level component to the north-west whose position angle, though in agreement with the jet axes of Tu08 and MJ2011, indicates activity in a direction previously associated with the receding jet. Variability of the north-western component, likely the result of re-energisation by unseen outflows, cannot be physically reconciled with the observed flare but if caused by an immediately previous periastron event implies a proper motion of 60 mas day$^{-1}$. Given the observed differences in jet behaviour compared to results prior to 2008, we believe that this work, in conjunction with results from Ca11 and our partnered submission, MJ2011, suggest Cir X-1's jet behaviour; including flow velocity and precession parameters, can vary on secular timescales.

\section{Acknowledgements}
D.E.C. and J.W.B. acknowledge support from the United Kingdom Science and Technology Facilities Council. The Australia Telescope Compact Array is part of the
Australia Telescope funded by the Commonwealth of Australia
for operation as a National Facility managed by CSIRO.


\begin{thebibliography}{}

\bibitem{} Brandt, W. N., Schulz, N. S., 2000, ApJ, 544, L123

\bibitem{} Calvelo, D. E., Fender, R. P., Broderick, J., Moin, A., Tingay, S., Tzioumis, A. K., Nicolson, G., 2010, ATel, 2699

\bibitem{} Corbel, S., Fender, R. P., Tzioumis, A. K., Tomsick, J. A., Orosz, J. A., Miller, J. M., Wijnands, R., Kaaret, P., 2002, Sci, 298, 196

\bibitem{} Dubner, G. M., Holdaway, M., Goss, W. M., Mirabel, I. F., 1998, AJ, 116, 1842

\bibitem{} Fender, R. P. \etal, 1998, ApJ, 506, L21

\bibitem{} Fender, R. P., Pooley, G. G., 2000, MNRAS, 318, L1

\bibitem{} Fender, R. P., Wu, K., Johnston, H., Tzioumis, A. K., Jonker, P., Spencer, R., van der Klis, M., 2004, Nature, 427, 222

\bibitem{} Fender, R. P., Tzioumis, A. K., Tudose, V., 2005, ATel, 563

\bibitem{} Fomalont, E. B., Geldzahler, B. J., Bradshaw, C. F., 2001, ApJ, 553L, 27

\bibitem{} Glass, I. S., 1978, MNRAS, 183, 335

\bibitem{} Goodall, P. T., Fathallah, A., Blundell, K.M., 2011, MNRAS, 529

\bibitem{} Haynes, R. F., Jauncey, D. L., Murdin, P. G., Goss, W. M., Longmore, A. J., Simons, L. W. J., Milne, D. K., Skellern, D. J., 1978, MNRAS, 185, 661

\bibitem{} Heinz, S., Schulz, N. S., Brandt, W. N., Galloway, D. K., 2007, ApJ, 663, L93

\bibitem{} Iaria, R., D'A\'{i}, A., Lavagetto, G., Di Salvo, T., Robba, N. R., Burderi, L., 2008, ApJ, 673, 1033

\bibitem{} Linares, M., \etal, 2010, ApJ, 719, L84

\bibitem{} Maccarone, T. J., 2002, MNRAS, 336, 1371

\bibitem{} Middelberg, E., Sault, R.J., Kesteven, M.J., 2006, PASA, 23, 147

\bibitem{} Nakajima, M., \etal, 2010, ATel, 2608

\bibitem{} Sault, R. J., Teuben, P. J., Wright, M. C. H., 1995, ASPC, 77, 433

\bibitem{} Shirley, R. E., Levine, A. M., Bradt, H. V., 1999, ApJ, 524, 1048

\bibitem{} Soleri, P., \etal, 2009, MNRAS, 397, L1 

\bibitem{} Stewart, R. T., Caswell, J. L., Haynes, R. F., Nelson, G. J., 1993, MNRAS, 261, 593

\bibitem{} Taylor, G. B., Carilli, C. L., Perley, R. A., 1999, ASPC, 180

\bibitem{} Tennant, A. F., Fabian, A. C., Shafer, R. A., 1986, MNRAS, 219, 871

\bibitem{} Thomas, R. M., Duldig, M. L., Haynes, R. F., Murdin, P., 1978, MNRAS, 185, P29

\bibitem{} Tudose, V., Fender, R. P., Kaiser, C. R., Tzioumis, A. K., van der Klis, M., Spencer, R. E., 2006, MNRAS, 372, 417 

\bibitem{} Tudose, V., Fender, R. P., Kaiser, C. R., Tzioumis, A. K., Spencer, R. E., van der Klis, M., 2008, MNRAS, 390, 447 [Tu08]

\bibitem{} Whelan, J. A. J., \etal, 1977, MNRAS, 181, 259 

\end{thebibliography}
\end{document}